# Efficiently Evacuating Lower Manhattan

**Team 21**

Caitlin Feltner (cmf269), Emily Lewis (eul4), Jamie Peck (jp2484), Mark Shipps (mas878), Scott Holmdahl (skh73)

**Abstract**

Although each year brings rapid advancement in meteorology and forecasting technology, the threat of natural disasters is not totally predictable. Predictability, however, still will not guarantee avoidability, thus adequate time to react to predictions remains a chief concern. Considerable effort by city authorities is often put forth to prepare for such events by investing in infrastructure and research on evacuation strategies to minimize the potential harm to humanity. This study focuses on the development of one of these evacuation strategies for Manhattan Island. The algorithm proposed is based on evaluating source-destination node routes with calculated traffic flow-densities and time costs. By utilizing a transshipment LP optimization model, we are able to understand the optimal throughput of evacuees in targeted exits of lower Manhattan to remove the population from harm's way in the least amount of time, assuming a potential natural disaster is forecasted.

**Introduction**

New York City (NYC) is the most populated city in the United States. As of July 1, 2018, the U.S. Census Bureau has appraised New York City's population at 8,398,748 (7). The city consists of 5 boroughs. Of the 5 boroughs, Manhattan is both the smallest and most densely populated (10). Manhattan alone has an estimated population of 1.63 million people (10). At only 23 square miles, Manhattan's population density is 70,826 people per square mile (10).

In the case of a potential natural disaster where emergency evacuations are ordered, 1.63 million civilians would be required to leave the borough at the same time. As a primary means of evacuation, vehicular transport leads to significant congestion. A severe case of this was in 2005 as millions of civilians fled a hurricane and got stuck in a 100-mile-long traffic jam, exacerbated by the many families' running out of gas while waiting in the congestion (1). Some intuitive strategies to help alleviate this congestion involve contraflow in evacuation routes to maximize the flow capacities of the road infrastructure, as Governor Christie ordered for Route 75 during the evacuation of Long Island for Hurricane Irene (2). This is a tremendous undertaking in itself, though, involving mass-coordination across multi-city authorities and resources. However, it is not just the challenge of vehicle congestion that makes an evacuation of Manhattan a difficult case. Over 75% of the households in Manhattan do not own a vehicle (4). These individuals would need to seek transportation out of the borough by alternate means such as bus, train, subway, ferry, or plane, placing an uncharacteristic stress on those systems as well.

Communication of the evacuation itself always plays an important role. With such a diverse contingency, evacuation orders of Manhattan must ensure that all common languages spoken are accounted for. In the case of an uncommon natural event for the area, such as a hurricane, the relative perception of the individuals must be considered. In the evacuation of New York City during Hurricane Irene, reports suggested citizens inexperienced with the situation were confused as to whether to take the threat seriously, opting to stock up on supplies and ride the storm out in their homes (2). In a best-case scenario, if the population heeds the evacuation orders, there is always still the autonomy of the individuals to factor in; directed routes for target locations may not be the chosen routes by those people when they do leave. This could occur as a result of other routes being perceived quicker, accidents, varying final destinations, and many other possibilities.

In an effort to tackle a piece of Manhattan's evacuation challenge, this study aims to optimize the amount of time it takes to evacuate a targeted region of Manhattan's during an emergency event as aforementioned.

**Background**

With a scope of exiting via the primary means of automobile, ferry, or train, the proposed optimization model will consider flow capacity of evacuation routes as defined by starting neighborhood and destination exit point. Because the lower Manhattan region houses the bulk of the population during the midday hours, the focus will be on evacuating neighborhoods south of $42_{nd}$ Street (5). Within lower Manhattan, 13 neighborhoods have been identified: Chelsea, Garment District, Murray Hill, Gramercy, Stuyvesant Tower, Greenwich Village, East Village, Soho, Little Italy, Lower East Side, Tribeca, China Town, and the Financial District. Each of these neighborhoods will represent the starting point for its anticipated number of evacuees (i.e., neighborhood population).

The exit point in this model will include tunnels, ferry stations, bridges, and train stations. Each exit point can evacuate a maximum amount of people, which is determined by the available space or resources (e.g., number of lanes, trains and seat capacities, length of tunnels). If an exit point's maximum capacity is reached, additional evacuees will incur a time penalty to simulate congestion.

To reach these exit points, evacuees may travel via foot, bicycle, or motor vehicles (i.e. bus, car, motorcycle). Since most of Manhattan's population is car-less and bicycles may not be readily available, only a portion of evacuees will be able to travel by these methods. It is assumed that those exiting via foot or bicycle cannot exit using the tunnels because tunnels are intended for motor vehicle use only while others exiting via motor vehicles cannot exit using the train station or ferry access points because these exits are intended for people who travel by foot or bicycle.

**Literature Review**

There is significant research done in the evacuation optimization space, with scholars considering a wide variety of modeling techniques and influencing factors. The study by Chen and Zhan in 2008 examines the efficiency of simultaneous (i.e., all residents evacuate at once) and staged evacuations (i.e., certain "zones" evacuate at different times) in a grid system, a ring road structure, and a real-life road system. The study concludes that although neither simultaneous nor staged evacuations are universally more efficient, a staged evacuation that evacuates non-adjacent zones is more effective in grid system (3).

Exploring both a maximum flow model and minimum-cost maximum flow model, Li, Zhang, and Wang optimized road-based evacuation across a set of traffic intersections in 2013. Although the analysis is limited to small set of traffic intersections, it concludes that it is more efficient to utilize multiple evacuation destinations than a single evacuation destination (6).

In a more recent (2018) study, Yan, Liu, and Song used the Cell Transmission Model (CTM) to optimize evacuation while considering "unfairness", such that individuals in higher risk areas (who are generally de-prioritized in order to achieve the maximum collective evacuation success) are given an appropriately higher "social fairness" weight. The paper concludes that although a "social fairness" weight does increase the evacuation rate for high risk areas, it decreases the collective evacuation rate (9).

Also in 2018, Pyakurel, Dempe, and Dhamala focused on the application of contraflow evacuation strategies as network flow problems. Their model addressed a wide variety of problems such as maximum flow, earliest flow arrival, quickest lex-maximum, and transshipment. By introducing partial contraflow techniques, the study concluded that partial contraflow may be a more optimized approach to full contraflow in that the case study was able to evacuate the maximum number of people while preserving capacity in identified routes or other emergency service uses (8).

**Problem Statement**

The optimization problem for the large-scale evacuation of midtown and lower Manhattan is formally defined in this section. Manhattan, the most densely populated county in the United

States of America, includes a wide array of transportation options. These options include cars, buses, cars, motorcycles, pedestrians, bicycles, and ferries.

For some evacuation routes, there are multiple transportation methods available, which further complicates the evacuation process. For example, the Brooklyn Bridge includes both car lanes, which can be occupied by buses, cars, and motorcycles, and walking lanes, which can be occupied by pedestrians and bicycles. This necessitates "sub-optimization" problems, as each potential evacuation route (e.g., bridge, tunnel) must be optimized across a set of transportation methods.

The objective is to minimize the total time to evacuate the entire population by allocating people to different evacuation routes, thereby providing an optimal rapid evacuation strategy.
The relevant parameters in the optimization model are summarized as follows:
- The number of neighborhoods (starting points)
- The number of exits
- The population of each neighborhood
- The capacity of an exit route via bus, car, and motorcycle on bridge
    - The number of lanes available
    - The length of the bridge
    - The quantity of buses available
    - The quantity (or ratio) of car owners
    - The quantity (or ratio) of motorcycle owners
    - The capacity of buses
    - The capacity of cars
    - The speed of buses
    - The speed of cars
    - The speed of motorcycles
- The capacity of an exit route via bus and car in tunnel
    - The number of lanes available
    - The length of the tunnel
    - The quantity of buses available

- The quantity (or ratio) of car owners
- The quantity (or ratio) of motorcycle owners
- The capacity of buses
- The capacity of cars
- The speed of buses
- The speed of cars
- The speed of motorcycles
- The capacity of an exit route via walking and biking on bridge
    - The number of sidewalks available
    - The length of the bridge
    - The quantity (or ratio) of bicycle owners
    - The speed of pedestrians
    - The speed of bicycles
- The capacity of an exit route via ferry stops
    - The number of available ferries
    - The ferry passenger capacity
- The transportation time to an exit route
    - The distance from starting points (e.g., neighborhood) to exit routes (e.g., bridge)

Major decision variables:

- The number of people evacuated to an evacuation route
- The discrete selection of transportation (by foot, bicycle, or motor vehicle) for each evacuation route

Constraints:

- The number of people evacuated is equal to the total population (fully evacuated)
- If the number of people evacuated to an exit route exceeds its capacity, additional evacuees will incur a time penalty
- The number of people evacuated by a particular method of transportation does not exceed the quantity of individuals who have that method of transportation available (e.g., can't have more cars than car owners)

Assumptions:

- The population counts by neighborhood taken by the US Census Bureau are representative of the number of people in the area on a given day
- Resources will be used for outbound travel regardless of function during normal operations (i.e., inbound lanes convert to outbound lanes)
- All available resources will be used regardless of typical usage during normal operations (e.g., entire fleet of ferries available for evacuation)
- The entire population is able to utilize all possible methods of transportation (e.g., bicycle) because, for example, disabled individuals would be prioritized for appropriate methods of transportation (e.g., train)
- Intermediate trips are not considered
- Evacuees cannot change their evacuation after departure
- No new transportation entering the city (i.e., only outbound)
- The analysis is limited to lower Manhattan, as this area has the largest population density and is therefore both most likely to be difficult to evacuate under emergency conditions
- Identified exit routes are major bridges, tunnels, train stations, ferry stops that lead out from the borough
- Evacuees possess perfect knowledge of the evacuation process and adhere to their determined routes (super-user)

These assumptions enable the team to 1) model a complex and nuanced problem using quantitative methods; 2) conform with assumptions in published literature; 3) perform the computational analysis in a reasonable time frame.

**Model Formulation and Solution Algorithm**

In the following section, the general approach and formulas to calculate the associated parameters and constraints are explained. This includes the evacuation duration minimization, route times, time penalties, exit capacities, and resource accessibility. Figure 1 and 2 are shown below to clarify the referenced nomenclature.

**Figure 1:** *Pre-Model Derivations Nomenclature*

*Route Times $[t_{ijkw}]$ Derivation:*
- $d(i,j)$: distance from neighborhood i to exit j
- $s_{walk}$: the time it takes to walk one mile

- $s_{bike}$: the time it takes to bike one mile
- $s_{drive}$: the time it takes to drive one mile
- $c_{car}(i,j)$: the congestion parameter for vehicle traffic for each route
    - $int(i,j)$: the approximate number of intersections on a route
    - $LOS$: time measure of traffic quality at each intersection
- $tp(j,k)$: time penalty for exit j via transportation mode k

*Time Penalties [$tp(j,k)$] Derivation:*
- For Bridges and Tunnels
    - $length(j)$: the length of exit j
    - $s_{walk}$: the time it takes to walk one mile
    - $s_{bike}$: the time it takes to bike one mile
    - $s_{drive}$: the time it takes to drive one mile
- For Ferries
    - $length(j)$: the length of the ferry route for exit j and back
    - $s_{ferry}$: the time it takes to boat one mile
    - $load_{ferry}$: the time associated with on/off boarding
- For Trains
    - $s_{train}$: the average time to reach the next train stop outside of Lower Manhattan and back
    - $load_{train}$: the time associated with on/off boarding

*Maximum Capacities [$maxcap(j,k)$] Derivation:*
- For Bridges and Tunnels
    - $length(j)$: the length of exit j
    - $lane_{walk}(j)$: the number of walking lanes
    - $lane_{bike}(j)$: the number of bicycle lanes
    - $lane_{drive}(j)$: the number of vehicle lanes
    - $l_{person}$: the average length of a person
    - $l_{bike}$: the average length of a bike
    - $l_{car}$: the average length of a car
- For Ferries
    - $st_{ferry}$: the number of available seats per ferry
    - $n_{ferry}(j)$: the number of ferry boats at exit j
- For Trains
    - $st_{train}$: the number of available seats per train
    - $n_{train}$: the number of trains

**Figure 2:** *Model Nomenclature*

*Indices*
- $i$: set of starting points (neighborhoods) indexed by i
- $j$: set of evacuation exits indexed by j
- $k$: set of transportation modes indexed by k
- $w$: set of periods that incur exponentially increasing time penalties indexed by w

*Decision Variables*
- $x_{ijkw}$: the number of people from neighborhood hood i who take the route to exit j via mode k in wave w

*Parameters*
- $t_{ijkw}$: the time from neighborhood i to exit j via mode k in wave w*
- $maxcap(j,k)$: maximum capacity of exit j for mode k
- $p(i)$: the population of neighborhood i

The approach that used to solve this optimization is a variation of the transshipment problem. Each neighborhood of Lower Manhattan is shown in red as a point source node. Each of the evacuation exits are shown in blue as a point destination or exit node. The routes that connect the source and destination nodes vary by evacuee capacity, distance to travel, and mode of transportation allowed. The model assumes all evacuees will start evacuating at the same time, and if a particular route reaches its nominal capacity, additional evacuees must incur a time penalty to simulate the effects of congestion.

**Figure 3:** *Model Approach*

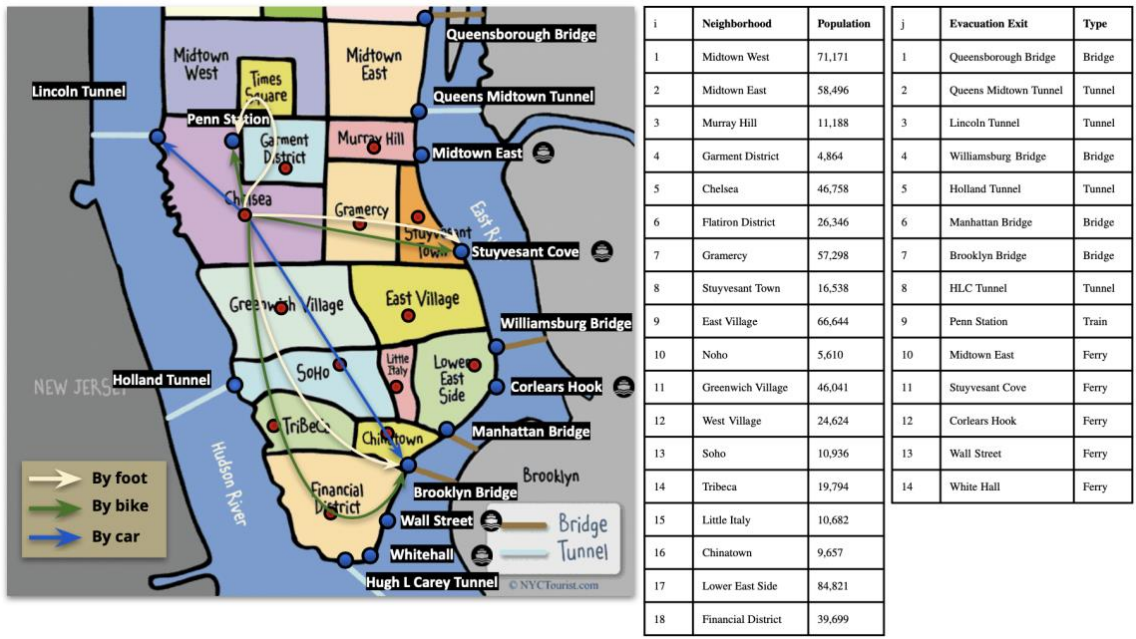

| i | Neighborhood | Population | j | Evacuation Exit | Type |
|---|---|---|---|---|---|
| 1 | Midtown West | 71,171 | 1 | Queensborough Bridge | Bridge |
| 2 | Midtown East | 58,496 | 2 | Queens Midtown Tunnel | Tunnel |
| 3 | Murray Hill | 11,188 | 3 | Lincoln Tunnel | Tunnel |
| 4 | Garment District | 4,864 | 4 | Williamsburg Bridge | Bridge |
| 5 | Chelsea | 46,758 | 5 | Holland Tunnel | Tunnel |
| 6 | Flatiron District | 26,346 | 6 | Manhattan Bridge | Bridge |
| 7 | Gramercy | 57,298 | 7 | Brooklyn Bridge | Bridge |
| 8 | Stuyvesant Town | 16,538 | 8 | HLC Tunnel | Tunnel |
| 9 | East Village | 66,644 | 9 | Penn Station | Train |
| 10 | Noho | 5,610 | 10 | Midtown East | Ferry |
| 11 | Greenwich Village | 46,041 | 11 | Stuyvesant Cove | Ferry |
| 12 | West Village | 24,624 | 12 | Corlears Hook | Ferry |
| 13 | Soho | 10,936 | 13 | Wall Street | Ferry |
| 14 | Tribeca | 19,794 | 14 | White Hall | Ferry |
| 15 | Little Italy | 10,682 | | | |
| 16 | Chinatown | 9,657 | | | |
| 17 | Lower East Side | 84,821 | | | |
| 18 | Financial District | 39,699 | | | |

*Objective Function*

The objective function (1) aims to minimize the aggregated duration of evacuation, which is represented by the summation of the durations and number of evacuees for chosen routes. In total, the model optimizes across 5,292 decision variables. Since the model assumes a total evacuation where everyone leaves at once, the minimum time to evacuate all of Lower Manhattan is calculated by taking the maximum of the evacuation durations.

$$(1)\ minimize\ \sum_{i=1}^{18}\sum_{j=1}^{14}\sum_{k=1}^{3}\sum_{w=1}^{7} t_{ijkw} * x_{ijkw}$$

Time ($t_{ijkw}$) in this objective function is defined as (distance of route * speed of transportation) + congestion parameter, where the congestion parameter is applicable to vehicular routes only.

$$(a)\ t_{ijkw} = d(i,j) * s_{walk}, k = 1\ and\ \forall i,j$$
$$(b)\ t_{ijkw} = d(i,j) * s_{bike}, k = 2\ and\ \forall i,j$$
$$(c)\ t_{ijkw} = d(i,j) * s_{drive} + c_{car}, k = 3\ and\ \forall i,j$$

In order to understand how the time parameter is calculated, we must first examine the distance between a starting neighborhood i and a possible exit j. This will define the route distance, and we calculated this based on Google Maps data. The speeds for each mode of transportation were derived from average values of walking, biking, and driving the distance of one mile (17). This data is easily obtainable with a web search. The congestion parameter ($c_{car}$) is a technique used to help simulate route congestion in high-traffic. A common measure of representing traffic flows in a route is level-of-service, which is a qualitative explanation of how quickly vehicles travel in a route, broken into typically six lettered tiers. Based on a Level-of-Service rating of E, the routes in this model all are assumed to be operating at capacity with unsteady traffic flow. This yields a 55-80 second time delay for every signaled intersection (12). Using Google Maps, we were able to count all signaled intersections within each route to derive the associated congestion parameter of that route.

$$(d)\ c_{car} = LOS * int(i,j)$$

An additional nuance in the calculation of time that is not shown in the aforementioned time equation is the concept of a time penalty ($tp(j,k)$) for saturated exits. Similar to the congestion parameter, this, too, helps to simulate time delays due to evacuees using exits that are already at capacity. Each type of exit was treated uniquely when deriving this time penalty. For example, bridge and tunnel time penalties were derived by taking the length of the bridge or tunnel, then using the speed of each type of transportation mode to compute a time that would represent how long it would take for an evacuee to "clear" that particular exit. The rationale for this was that additional evacuees wanting to use the exit would first have to wait until another evacuee has cleared.

$$(e)\ tp(j,k) = length(j) * s_{walk}, j = 1,4,6,7\ and\ k = 1$$
$$(f)\ tp(j,k) = length(j) * s_{bike}, j = 1,4,6,7\ and\ k = 2$$
$$(g)\ tp(j,k) = length(j) * s_{drive}, j = 1,\ldots 8\ and\ k = 3$$

For ferry and train exit types, the time penalty was derived by taking the distance of the particular ferry or route round-trip and its speed, then adding an additional time delay to account for the ferry's unloading and loading of passengers. The speeds and distances for these calculations were all found by Google Maps and web search data.

$$(h)\ tp(j,k) = length(j) * s_{ferry} + load_{ferry}, j = 10,11,12,13,14\ and\ k = 1,2$$
$$(i)\ tp(j,k) = length(j) * s_{train} + load_{train}, j = 9\ and\ k = 1,2$$

To incorporate this time penalty concept into the time equation, we opted to treat a penalized route as a completely separate route with its own distance (remaining the same as the original) and associated time-to-travel (which reflects the route's original time plus the added time penalty). If a given route in the model reaches capacity and incurs a time penalty, its original route index is essentially no longer available. Furthermore, to compound the effects of congestion in an exit, the time penalty is increased at an exponential rate when a particular exit reaches a defined capacity. This notion of capacity thresholds is called "waves" and includes up to 7 of these overflow waves for each route to ensure model feasibility.

$$(j)\ tp(j,k), w = 1;\ 2^{w-2} * tp(j,k), w = 2,\ldots 7$$

**Figure 4:** *Wave Formulation*

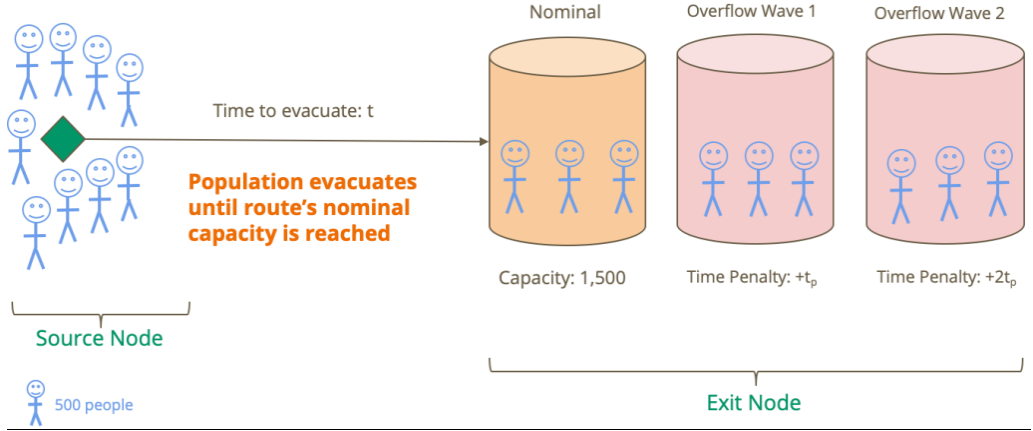

Each wave's defined capacity is based on the exit's nominal capacity. The nominal capacities for bridges and tunnels were calculated by taking the length of the exit, dividing it by the average length a person, bike, or car would occupy, and multiplying by the respective number of pedestrian, bike, and car lanes.

$$(k) maxcap(j,k) = length(j)/l_{person} * lane_{walk}(j), j = 1, 4, 6, 7 \text{ and } k = 1$$
$$(l) maxcap(j,k) = length(j)/l_{bike} * lane_{bike}(j), j = 1, 4, 6, 7 \text{ and } k = 2$$
$$(m) maxcap(j,k) = length(j)/l_{car} * lane_{drive}(j), j = 1, \ldots 8 \text{ and } k = 3$$

For ferries and train stations, the capacities were derived by multiplying the number of seats available by the number of ferries and trains available at the station and associated ports.

$$(n) maxcap(j,k) = st_{ferry} * n_{ferry}, j = 10, 11, 12, 13, 14 \text{ and } k = 1,2$$
$$(o) maxcap(j,k) = st_{train} * n_{train}, j = 9 \text{ and } k = 1,2$$

*Constraints*

The explicit formulations for the constraints are found in the following subsections. Each subsection represents a category of constraints applied in the model.

*Population Constraint*

Constraint (2) imposes a complete evacuation in which all persons accounted for in each neighborhood must leave Lower Manhattan. The total number of evacuated people from a given neighborhood can be calculated as the summation of routed evacuees across all feasible exit points, transportation modes, and waves (13).

(2) $\sum_{j=1}^{14} \sum_{k=1}^{3} \sum_{w=1}^{7} x_{ijkw} = p(i), \quad \forall i$

*Maximum Capacity Constraints*

The model assumes that a given exit capacity can vary depending on the type of transportation evacuees take to arrive there. For example, major bridges contain vehicle lanes, pedestrian walkways, and bike lanes. Each of these crossover means can carry different amounts of people depending on the structural dimensions. Constraints (3) to (5) ensures that each exit does not reach over capacity for those that traveled by foot ($k = 1$), by bike ($k = 2$), and by car ($k = 3$), respectively. A special case is considered for evacuees who walk or bike to ferry port exits. Since there is no structural delineation between these two types of evacuees on a ferry, constraint (6) states that the combined number of walkers and bicyclists cannot be greater than the exit's maximum capacity.

(3) $\sum_{i=1}^{18} x_{ij1w} \leq maxcap(j,k), \quad for\ w = 1,2,\dots 7$

(4) $\sum_{i=1}^{18} x_{ij2w} \leq maxcap(j,k), \quad for\ w = 1,2,\dots 7$

(5) $\sum_{i=1}^{18} x_{ij3w} \leq maxcap(j,k), \quad for\ w = 1,2,\dots 7$

(6) $\sum_{i=1}^{18} \sum_{k=1}^{2} x_{ijkw} \leq maxcap(j,k), \quad for\ j = 10,11,\dots 14\ and\ w = 1,2,\dots 7$

*Infeasible Route Constraints*

The model accounted for associated safety risks based on the type of exit for each transportation mode. Based on its posed danger, infeasible routes were identified and implemented so that no person can evacuate in that way. For example, since tunnels do not provide adequate visibility or space for non-motor vehicles, constraint (7) ensures that evacuees who travel by foot or bike cannot exit through a tunnel. Driving to ferry ports and train stations was assumed to worsen congestion and increase general confusion as evacuees try to find parking prior to boarding. Therefore, constraints (8) and (9) enforces that evacuees who travel by car can neither exit by ferry ports or train stations, respectively.

(7) $\sum_{i=1}^{18} \sum_{k=1}^{2} x_{ijkw} = 0, \quad for\ j = 2,3,5,8\ and\ w = 1,2,\dots 7$

(8) $\sum_{i=1}^{18} x_{ij1w} = 0, \quad for\ j = 10,11,\dots 14\ and\ w = 1,2,\dots 7$

(9) $\sum_{i=1}^{18} x_{ij1w} = 0, \quad for\ j = 9\ and\ w = 1,2,\dots 7$

*Transportation Accessibility Constraints*

Considering that Manhattan heavily relies on its public transportation systems, the model accounted for the accessibility to personal means of transportation. This was derived from general statistics on vehicle ownership and bike usage in Manhattan. It is estimated that 22% of Manhattan residents own a car while 24% ride a bike at least once every year. As such, constraint (10) states that up to 24% of evacuees can bike while constraint (11) states that up to 22% of evacuees can drive to their exits (14, 15). Since it is uncertain where bicyclists and car owners reside, both constraints are applied to each neighborhood rather than the population as a whole.

$$(10) \sum_{j=1}^{14} \sum_{w=1}^{7} x_{ij2w} \leq 0.24 * p(i)$$

$$(11) \sum_{j=1}^{14} \sum_{w=1}^{7} x_{ij3w} \leq 0.22 * p(i)$$

Because the model is constructed as a Linear Programming (LP) problem, the outputs for decision variables include non-integers. In reality, people must be quantified as integer values. This could be addressed by reformulating the model as a Multi-Integer Linear Programming problem where each decision variable is modeled as a general integer. However, the inclusion of integer variables makes an optimization much more difficult to solve. Often times, the memory and solution time rises exponentially as more integer variables are considered and solvers prove to be extremely sensitive to the formulation (16). As a result, the model retained an LP formulation and employed the CPLEX solver, in which its core solution is the Simplex Algorithm. The optimal solution to our model was found in 0.01 seconds. While other LP solvers are adequate (Gurobi, CONOPT, etc.), we chose CPLEX as the model solver.

**Results and Discussion**

Our model indicated that the total evacuation time of Lower Manhattan was 3 hours and 19 minutes. The maximum amount of bikes and cars were utilized in this model, and a sample of the model results can be seen in Table 1 below. This sample of results shows the starting districts of Chelsea, Chinatown and East Village and illustrates a sample of each starting point and possible ending point with the respective number of travelers categorized by transportation type. The complete results can be seen in the Supplementary Materials section of this report.

**Table 1**: *Results Snapshot*

| Starting | Exit Point | By Foot | By Bike | By Car | Sum of Total |
|---|---|---|---|---|---|

| District | | | | | |
|---|---|---|---|---|---|
| **Chelsea** | **Total of All Exits** | **25249.32** | **11221.92** | **10286.76** | **46758** |
| Chelsea | Brooklyn Bridge | 23952 | 0 | 2473.06 | 26425.06 |
| Chelsea | Holland Tunnel | 0 | 0 | 7813.7 | 7813.7 |
| Chelsea | Penn Station | 0 | 11221.92 | 0 | 11221.92 |
| Chelsea | Whitehall | 1297.32 | 0 | 0 | 1297.32 |
| **Chinatown** | **Total of All Exits** | **5214.78** | **2317.68** | **2124.54** | **9657** |
| Chinatown | Manhattan Bridge | 5214.78 | 2317.68 | 2124.54 | 9657 |
| **East Village** | **Total of All Exits** | **35987.76** | **15994.56** | **14661.68** | **66644** |
| East Village | Manhattan Bridge | 3055.26 | 0 | 0 | 3055.26 |
| East Village | Queensborough Bridge | 0 | 15994.56 | 0 | 15994.56 |
| East Village | Wall Street | 12722.76 | 0 | 0 | 12722.76 |
| East Village | Williamsburg Bridge | 20209.74 | 0 | 14661.68 | 34871.42 |

To provide a visual of the overall results, Figure 5 shows below each starting district and its corresponding evacuation exit by quantity.

**Figure 5:** *Results Visual by Population Density, Starting District, and Exit*

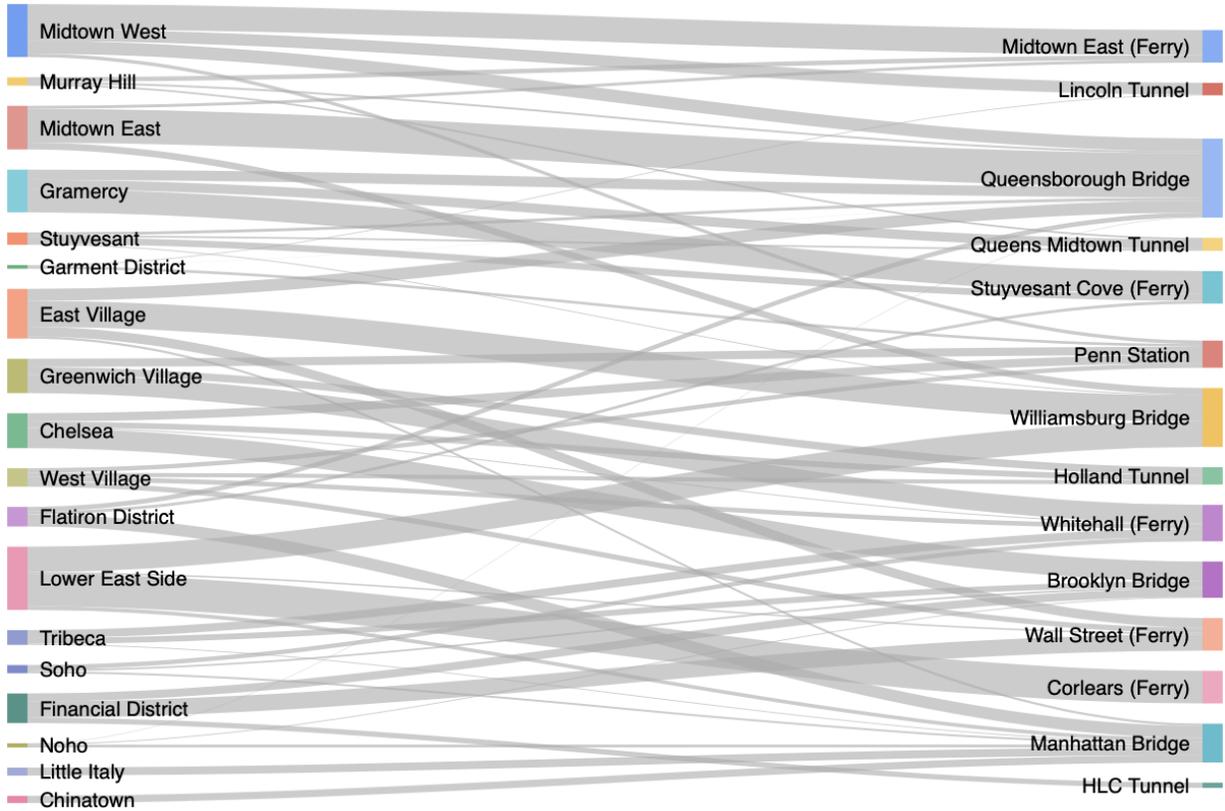

Based on the results, it is clear that districts that have a larger population, a larger area, and access to all types of transportation use the maximum amount of evacuation routes, whereas less populated, smaller area districts use minimal amounts of evacuation routes. For example, East Village, Chelsea, Lower East Side, Midtown West, Stuyvesant, and West Village all use four evacuation routes for their respective districts, which is the maximum amount of evacuation routes we see used for one district. Contrarily, Chinatown and Little Italy both use only one evacuation route for their respective districts.

The results also accurately reflect the assumptions and constraints that the model used for transportation method breakdown. Table 2 confirms this, showing that with the results produced, the constraint matches the percent of population that utilized bikes and cars to evacuate.

Table 2: *Transportation Method Breakdown*

| Transportation Method | Quantity | Percent of Population | Constraint |
|---|---|---|---|
| Walking | 330,030 | 54% | N/A |
| Bike | 146,680 | 24% | No more than 24% |
| Car | 134,456 | 22% | No more than 22% |
| *Total* | *611,167* | *100%* | *N/A* |

The final analysis completed via the model results is the breakdown of exit point utilization for each distinct district. Table 3 below shows the breakdown of the Chelsea exit point utilizations, which is one of the most complex evacuation strategies for a district in Lower Manhattan. Chelsea utilizes four different exit points: Brooklyn Bridge, the Holland Tunnel, Penn Station, and Whitehall, and shows that most of the population exits through the Brooklyn Bridge and Penn Station, with Holland Tunnel and Whitehall exit points only being utilized as supporting strategies.

Table 3: *Chelsea Exit Point Utilization*

| **Exit Point** | **Exit Point Utilization** |
|---|---|
| Brooklyn Bridge | 56.5% |
| Holland Tunnel | 16.7% |
| Penn Station | 24% |
| Whitehall | 2.8% |

**Conclusion and Recommendations**

Based on the results of the model, a series of conclusions on how to evacuate Lower Manhattan efficiently were made.

First, the model indicates that the maximum possible quantity of bicycles and cars should be utilized during evacuation, as both bicycles and cars were utilized in our optimized evacuation at values which correspond to their respective constraints: 24% and 22%. This indicates that any individual with a bicycle or car in Lower Manhattan should utilize this faster mode of transportation because it reduces the overall evacuation duration. It's worth noting that this result probably stems from our model's inability to effectively account for the subtle inefficiencies associated with congestion effects, such as New York City streets having limited capacity for the millions of individuals evacuating Lower Manhattan.

Another significant takeaway from our model is that 100% of individuals residing in Little Italy and Chinatown should evacuate over the Manhattan Bridge. This makes sense because, based on the relatively small size of Little Italy and Chinatown and the ability of the Manhattan Bridge to accommodate a large quantity of evacuees, this is probably the most efficient evacuation route for all residents of Little Italy and Chinatown.

*Figure 5: Chinatown Exit Routes*

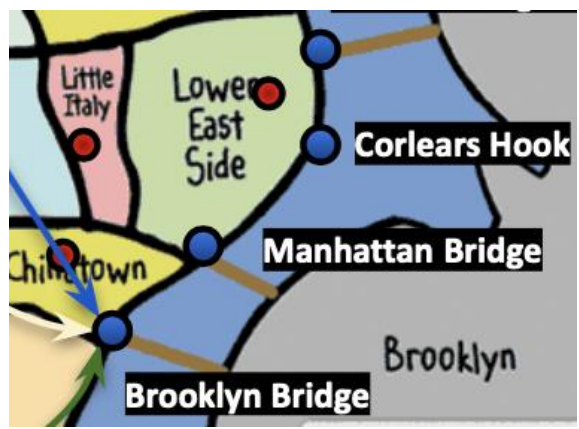

Additionally, we can look at particular districts which have a distributed set of evacuation routes and note that Chelsea, East Village, Lower East Side, Midtown West, West Village, and Stuyvesant all utilize four evacuation routes, which represents the largest quantity of evacuation routes utilized by any given neighborhood. This indicates that these neighborhoods are probably the most complex neighborhoods to evacuate, so New York City should encourage special evacuation preparation and devote additional evacuation resources to these districts in order to efficiently evacuate these neighborhoods. This result aligns with intuition, as residents of those districts have a multitude of potential evacuation options available.

To improve the accuracy and validity of our model, some potential improvements include modelling uncertainty, accounting for population "pulsing", incorporating congestion data, considering other evacuation methods, modelling evacuation staging, increasing source node fidelity, routing traffic along New York City's grid system, and incorporating imperfect decision-making.

Uncertainty in our model would allow the evacuation strategy to account for parameters which might vary on a day-by-day basis, such as the speed of each transportation methods. The rate which a given transportation method travels may vary depending on the weather because, for example, an evacuating individual can ride a bike much faster on a sunny day than a snowy day. For Lower Manhattan, the quantity of individuals in the area at any given time is not fully represented by the population as the area experiences "pulsing" in the sense that many New York City residents commute into Lower Manhattan from other areas of the city and the population increases dramatically during the daytime due to tourists visiting the area. Our model does not account for this pulsing effect. This could be modelled applying uncertainty to the total population, but this may cause the model to just optimize for the maximum possible population (roughly 4 million, which occurs during the peak of the "pulse") instead of the residential population of Lower Manhattan (roughly 600 thousand). Instead, one could develop two models: one for the residential population of Lower Manhattan and one for the peak population due to pulsing.

Another potential improvement to the model would be developing a data-driven model for route flow rates and congestion effects. To illustrate this, we can consider two extremes: when there are very few cars on the road, and when there are many cars on the road. When the roads are relatively open with few cars on the road, the cars that are on the road should experience little to no traffic and therefore travel near their maximum speed. In contrast, when there are many cars on the road, there will be significant congestion, especially at key evacuation locations like ferry ports. The challenge is modelling these congestion effects and, unfortunately, the team was unable to develop an effective method for modelling them, but it would be an interesting and significant factor to explore further.

Two other potential factors to incorporate into our model are other modes of evacuation and evacuation staging. By adding other potential evacuation methods, such as shuttle buses and helicopters, the model could more accurately account for the effects of individuals utilizing these evacuation methods. Additionally, the model could include evacuation stages, which involves the orderly withdrawal of certain subsets of the population in different stages. It's unclear how much impact these factors would have on the model, as additional transportation methods could be considered edge cases and evacuation staging would be very difficult to implement in real life.

A straightforward method of improving the model is increasing source node fidelity. The current model considers each residential neighborhood to be a geographical point where thousands of individuals begin their evacuation. In reality, individuals are distributed throughout each neighborhood, but it would be difficult to effectively model distance between locations when neighborhoods are considered a distributed set of locations rather than a single point.

Another potential method for improving the model would be to incorporate New York City's grid system as a factor in the model. For example, if a particular group needs to travel two blocks north or two blocks west, there is only one efficient route: directly horizontal or direct vertical. However, if that group needs to travel one block north and one block west, there are two potential routes: one block north then one block west, or one block west then one block north. To account for this de-congestion by splitting traffic through different, equivalent routes on New York City's grid system, the model could reduce the time penalties of congestion effects for journeys with many possible routes.

The final method for improving our model would modelling human's making imperfect decisions. For example, it may be optimal for everyone in a particular neighborhood by walking towards a particular route, but individuals in that area may choose to evacuate by car because they are attached to their car or they may choose a suboptimal evacuation route because they have family at the end of the suboptimal route. This could potentially be modelled by using uncertainty by including a constraint that no more than 90% will choose the optimal route, or it could be modelled by adding an additional variable to account for individuals making imperfect decisions.

In conclusion, despite some limitations to our model, it illustrates numerous significant results related to evacuation strategy for Lower Manhattan. A full-fledged model tackling a similar problem could be leveraged by New York City to effectively evacuate residents of its most densely-populated neighborhoods in the most time-efficient method possible, which could save lives and allow Lower Manhattan residents to feel comfortable in the event of a forecasted emergency situation.

**Supplementary Materials**

1. evacuation_model.gms
2. Evacuation_data.xlsx

# Appendix

*Appendix A - Results*

| Row Labels | Sum of by foot | Sum of by bike | Sum of by car | Sum of total |
|---|---|---|---|---|
| **chelsea** | **25249.32** | **11221.92** | **10286.76** | **46758** |
|   brooklyn-bridge | 23952 | 0 | 2473.06 | 26425.06 |
|   holland-tunnel | 0 | 0 | 7813.7 | 7813.7 |
|   penn-station | 0 | 11221.92 | 0 | 11221.92 |
|   whitehall | 1297.32 | 0 | 0 | 1297.32 |
| **chinatown** | **5214.78** | **2317.68** | **2124.54** | **9657** |
|   manhattan-bridge | 5214.78 | 2317.68 | 2124.54 | 9657 |
| **east-village** | **35987.76** | **15994.56** | **14661.68** | **66644** |
|   manhattan-bridge | 3055.26 | 0 | 0 | 3055.26 |
|   queensborough-bridge | 0 | 15994.56 | 0 | 15994.56 |
|   wall-street | 12722.76 | 0 | 0 | 12722.76 |
|   Williamsburg-bridge | 20209.74 | 0 | 14661.68 | 34871.42 |
| **financial-district** | **21437.46** | **9527.76** | **8733.78** | **39699** |
|   brooklyn-bridge | 0 | 9041.94 | 1742.14 | 10784.08 |
|   hlc-tunnel | 0 | 0 | 6991.64 | 6991.64 |
|   wall-street | 21437.46 | 485.82 | 0 | 21923.28 |
| **flatiron-district** | **14226.84** | **6323.04** | **5796.12** | **26346** |
|   manhattan-bridge | 10348.28 | 0 | 5796.12 | 16144.4 |
|   queensborough-bridge | 0 | 6323.04 | 0 | 6323.04 |
|   Stuyvesant-cove | 3878.56 | 0 | 0 | 3878.56 |
| **garment-district** | **2626.56** | **1167.36** | **1070.08** | **4864** |
| Lincoln-tunnel | 0 | 0 | 1070.08 | 1070.08 |
| penn-station | 2626.56 | 1002.48 | 0 | 3629.04 |
| queensborough-bridge | 0 | 164.88 | 0 | 164.88 |
| **gramercy** | **30940.92** | **13751.52** | **12605.56** | **57298** |
| queens-midtown-tunnel | 0 | 0 | 12605.56 | 12605.56 |
| queensborough-bridge | 0 | 13751.52 | 0 | 13751.52 |
| Stuyvesant-cove | 30940.92 | 0 | 0 | 30940.92 |
| **Greenwich-village** | **24862.14** | **11049.84** | **10129.02** | **46041** |
| holland-tunnel | 0 | 0 | 10129.02 | 10129.02 |
| penn-station | 0 | 11049.84 | 0 | 11049.84 |
| whitehall | 24862.14 | 0 | 0 | 24862.14 |
| **little-Italy** | **5768.28** | **2563.68** | **2350.04** | **10682** |
| manhattan-bridge | 5768.28 | 2563.68 | 2350.04 | 10682 |
| **lower-east-side** | **45803.34** | **20357.04** | **18660.62** | **84821** |
| corlears-hook | 43750 | 0 | 0 | 43750 |
| manhattan-bridge | 0 | 4973.04 | 0 | 4973.04 |
| wall-street | 2053.34 | 0 | 0 | 2053.34 |
| Williamsburg-bridge | 0 | 15384 | 18660.62 | 34044.62 |
| **midtown-east** | **31587.84** | **14039.04** | **12869.12** | **58496** |
| midtown-east | 3945.58 | 0 | 0 | 3945.58 |
| queensborough-bridge | 18620 | 14039.04 | 12869.12 | 45528.16 |
| Williamsburg-bridge | 9022.26 | 0 | 0 | 9022.26 |
| **midtown-west** | **38432.34** | **17081.04** | **15657.62** | **71171** |
| Lincoln-tunnel | 0 | 0 | 15657.62 | 15657.62 |
| midtown-east | 33762.9 | 0 | 0 | 33762.9 |
| penn-station | 4669.44 | 0 | 0 | 4669.44 |
| queensborough-bridge | 0 | 17081.04 | 0 | 17081.04 |
| **Murray-hill** | **6041.52** | **2685.12** | **2461.36** | **11188** |
| midtown-east | 6041.52 | 0 | 0 | 6041.52 |
| queens-midtown-tunnel | 0 | 0 | 2461.36 | 2461.36 |
| queensborough-bridge | 0 | 2685.12 | 0 | 2685.12 |
| **NoHo** | **3029.4** | **1346.4** | **1234.2** | **5610** |
| brooklyn-bridge | 0 | 0 | 1234.2 | 1234.2 |
| manhattan-bridge | 3029.4 | 752.46 | 0 | 3781.86 |
| queensborough-bridge | 0 | 593.94 | 0 | 593.94 |
| **SoHo** | **5905.44** | **2624.64** | **2405.92** | **10936** |
| brooklyn-bridge | 0 | 0 | 2405.92 | 2405.92 |
| manhattan-bridge | 0 | 2624.64 | 0 | 2624.64 |
| whitehall | 5905.44 | 0 | 0 | 5905.44 |
| **Stuyvesant-town** | **8930.52** | **3969.12** | **3638.36** | **16538** |
| queens-midtown-tunnel | 0 | 0 | 2373.08 | 2373.08 |
| queensborough-bridge | 0 | 3969.12 | 0 | 3969.12 |
| Stuyvesant-cove | 8930.52 | 0 | 0 | 8930.52 |
| Williamsburg-bridge | 0 | 0 | 1265.28 | 1265.28 |
| **tribeca** | **10688.76** | **4750.56** | **4354.68** | **19794** |
| brooklyn-bridge | 0 | 3558.06 | 4354.68 | 7912.74 |
| manhattan-bridge | 0 | 1192.5 | 0 | 1192.5 |
| whitehall | 10688.76 | 0 | 0 | 10688.76 |
| **west-village** | **13296.96** | **5909.76** | **5417.28** | **24624** |
| holland-tunnel | 0 | 0 | 5417.28 | 5417.28 |
| penn-station | 0 | 5909.76 | 0 | 5909.76 |
| wall-street | 7050.62 | 0 | 0 | 7050.62 |
| whitehall | 6246.34 | 0 | 0 | 6246.34 |
| **Grand Total** | **330030.18** | **146680.08** | **134456.74** | **611167** |

*Appendix B - GAMS Code*

Sets
   i starting point / midtown-west, midtown-east, Murray-hill, garment-district, chelsea, flatiron-district, gramercy, Stuyvesant-town, east-village, NoHo, Greenwich-village, west-village, SoHo, tribeca, little-Italy, chinatown, lower-east-side, financial-district /
   j exits / queensborough-bridge, queens-midtown-tunnel, Lincoln-tunnel, Williamsburg-bridge, holland-tunnel, manhattan-bridge, brooklyn-bridge, hlc-tunnel, penn-station, midtown-east, Stuyvesant-cove, corlears-hook, wall-street, whitehall /;

Parameters
   p(i) population of starting point i in cases
   /   midtown-west        71171
       midtown-east        58496
       Murray-hill         11188
       garment-district    4864
       chelsea             46758
       flatiron-district   26346
       gramercy            57298
       Stuyvesant-town     16538
       east-village        66644
       NoHo                5610
       Greenwich-village   46041
       west-village        24624
       SoHo                10936
       tribeca             19794
       little-Italy        10682
       chinatown           9657
       lower-east-side     84821
       financial-district  39699 /

   mf(j) max capacity at exit j by foot
   /   queensborough-bridge    3724

```
    queens-midtown-tunnel   0
    Lincoln-tunnel          0
    Williamsburg-bridge     7308
    holland-tunnel          0
    manhattan-bridge        6854
    brooklyn-bridge         5988
    hlc-tunnel              0
    penn-station            7296
    midtown-east            8750
    Stuyvesant-cove         8750
    corlears-hook           8750
    wall-street             8750
    whitehall               12250 /

mb(j) max capacity at exit j by bike
/   queensborough-bridge    1306
    queens-midtown-tunnel   0
    Lincoln-tunnel          0
    Williamsburg-bridge     2564
    holland-tunnel          0
    manhattan-bridge        2404
    brooklyn-bridge         2100
    hlc-tunnel              0
    penn-station            7296
    midtown-east            8750
    Stuyvesant-cove         8750
    corlears-hook           8750
    wall-street             8750
    whitehall               12250 /

mc(j) max capacity at exit j by car
/   queensborough-bridge    11385
    queens-midtown-tunnel   8720
    Lincoln-tunnel          16320
    Williamsburg-bridge     19880
    holland-tunnel          11680
    manhattan-bridge        16310
    brooklyn-bridge         12210
    hlc-tunnel              12400
    penn-station            0
    midtown-east            0
    Stuyvesant-cove         0
    corlears-hook           0
    wall-street             0
    whitehall               0 /
```

tpf(j)  time penalty at exit j by foot

/   queensborough-bridge    18.4
    queens-midtown-tunnel   0
    Lincoln-tunnel          0
    Williamsburg-bridge     36.11
    holland-tunnel          0
    manhattan-bridge        33.86
    brooklyn-bridge         29.58
    hlc-tunnel              0
    penn-station            280
    midtown-east            18.14
    Stuyvesant-cove         18.28
    corlears-hook           17.97
    wall-street             18.40
    whitehall               20.38 /

tpb(j)  time penalty at exit j by bike

/   queensborough-bridge    4.23
    queens-midtown-tunnel   0
    Lincoln-tunnel          0
    Williamsburg-bridge     8.3
    holland-tunnel          0
    manhattan-bridge        7.79
    brooklyn-bridge         6.80
    hlc-tunnel              0
    penn-station            28
    midtown-east            18.14
    Stuyvesant-cove         18.28
    corlears-hook           17.97
    wall-street             18.40
    whitehall               20.38 /

tpc(j)  time penalty at exit j by car

/   queensborough-bridge    21.16
    queens-midtown-tunnel   36.44
    Lincoln-tunnel          45.49

```
     Williamsburg-bridge   41.52
     holland-tunnel        48.80
     manhattan-bridge      38.94
     brooklyn-bridge       34.02
     hlc-tunnel            51.80
     penn-station           0
     midtown-east           0
     Stuyvesant-cove        0
     corlears-hook          0
     wall-street            0
     whitehall              0 /;
```

Table d(i,j) distance from starting point i to exit j

| | queensborough-bridge | queens-midtown-tunnel | Lincoln-tunnel | Williamsburg-bridge | holland-tunnel | manhattan-bridge | brooklyn-bridge | hlc-tunnel | penn-station | midtown-east | Stuyvesant-cove | corlears-hook | wall-street | whitehall |
|---|---|---|---|---|---|---|---|---|---|---|---|---|---|---|
| midtown-west | 2.3 | 2.2 | 0.9 | 6.7 | 2.9 | 4.3 | 4.1 | 4.7 | 1.1 | 2.2 | 3 | 4.6 | 4.9 | 4.8 |
| midtown-east | 0.8 | 1.4 | 2.1 | 1.9 | 3.3 | 3.8 | 3.8 | 4.6 | 1.6 | 1.3 | 1.9 | 3.8 | 4.6 | 4.6 |
| Murray-hill | 1.3 | 0.5 | 2 | 2.7 | 2.8 | 3.2 | 3.2 | 4 | 1.1 | 0.5 | 1.1 | 3.1 | 3.9 | 4 |
| garment-district | 2.3 | 1.5 | 1 | 3.5 | 2.4 | 3.7 | 3.4 | 4.1 | 0.3 | 1.5 | 2 | 3.9 | 4.2 | 4.1 |
| chelsea | 3.2 | 2.6 | 1.6 | 3.1 | 1.5 | 3.2 | 2.4 | 3.3 | 0.6 | 2 | 1.7 | 3.5 | 3.5 | 3.4 |
| flatiron-district | 2.6 | 1.7 | 2.3 | 2.4 | 1.7 | 2.4 | 2.3 | 3 | 0.9 | 1.4 | 1 | 2.7 | 3.2 | 3.1 |
| gramercy | 2.3 | 1.3 | 2.7 | 2.1 | 1.8 | 2.3 | 2.2 | 3.1 | 1.2 | 1.1 | 0.7 | 2.3 | 3 | 3.1 |
| Stuyvesant-town | 2.4 | 1.4 | 3.3 | 1.4 | 2.1 | 2.4 | 2.4 | 3 | 1.8 | 1.1 | 0.5 | 1.7 | 2.9 | 3 |
| east-village | 2.8 | 1.8 | 3.5 | 1.1 | 1.8 | 2.1 | 2.1 | 2.7 | 2.2 | 1.5 | 1 | 1.3 | 2.6 | 2.7 |
| NoHo | 3 | 2 | 3.1 | 1.5 | 1.1 | 1.7 | 1.4 | 2.3 | 1.8 | 1.8 | 1.4 | 1.8 | 2.3 | 2.3 |
| Greenwich-village | 3.6 | 2.6 | 2.7 | 2.1 | 0.8 | 2.2 | 1.8 | 2.5 | 1.3 | 2.3 | 2 | 2.4 | 2.6 | 2.5 |
| west-village | 3.7 | 2.7 | 2.5 | 2.2 | 0.8 | 2.3 | 1.9 | 2.5 | 1.3 | 2.5 | 2.1 | 2.5 | 2.6 | 2.6 |
| SoHo | 3.8 | 2.9 | 3.3 | 1.6 | 0.5 | 1.4 | 1.1 | 1.7 | 2.1 | 2.6 | 2.2 | 1.7 | 1.9 | 1.8 |
| tribeca | 4.5 | 3.5 | 3.8 | 2 | 0.8 | 1.5 | 0.6 | 1.2 | 2.6 | 3.2 | 2.8 | 1.8 | 1.3 | 1.2 |

|             |     |     |     |     |     |     |
|-------------|-----|-----|-----|-----|-----|-----|
| little-Italy | 3.9 | 2.8 | 3.8 | 1.3 | 1   | 0.9 |
| 0.8 | 1.7 | 2.4 | 2.5 | 2.1 | 1.3 | 1.7 | 1.8 |
| chinatown | 4.1 | 3 | 4 | 1.4 | 1.1 | 0.7 |
| 0.7 | 1.6 | 2.8 | 2.7 | 2.3 | 1.3 | 1.6 | 1.6 |
| lower-east-side | 3.9 | 2.7 | 4.4 | 0.5 | 1.7 | 1.5 |
| 1.3 | 2 | 3.1 | 2.5 | 1.7 | 0.4 | 1.8 | 2 |
| financial-district | 4.8 | 3.8 | 4.3 | 2 | 1.5 | 1.6 |
| 0.5 | 0.7 | 3.3 | 3.5 | 3.1 | 1.7 | 0.6 | 0.7 |

 ;

Scalar f walking speed /20/;
Scalar b biking speed /6/;
Scalar c driving speed /45/;

***evacuation durations
** no time penalty
Parameter fc(i,j) time cost for walking;
   fc(i,j)=f*d(i,j);

Parameter bc(i,j) time cost for biking;
   bc(i,j)=b*d(i,j);

Parameter cc(i,j) time cost for driving;
   cc(i,j)=c*d(i,j);

** time penalty 1
Parameter fctp(i,j) time cost for walking;
   fctp(i,j)=f*d(i,j)+tpf(j);

Parameter bctp(i,j) time cost for biking;
   bctp(i,j)=b*d(i,j)+tpb(j);

Parameter cctp(i,j) time cost for driving;
   cctp(i,j)=c*d(i,j)+tpc(j);

** time penalty 2
Parameter fctp1(i,j) time cost for walking;
   fctp1(i,j)=f*d(i,j)+2*tpf(j);

Parameter bctp1(i,j) time cost for biking;
   bctp1(i,j)=b*d(i,j)+2*tpb(j);

Parameter cctp1(i,j) time cost for driving;
   cctp1(i,j)=c*d(i,j)+2*tpc(j);

** time penalty 3

Parameter fctp2(i,j) time cost for walking;
   fctp2(i,j)=f*d(i,j)+4*tpf(j);

Parameter bctp2(i,j) time cost for biking;
   bctp2(i,j)=b*d(i,j)+4*tpb(j);

Parameter cctp2(i,j) time cost for driving;
   cctp2(i,j)=c*d(i,j)+4*tpc(j);

** time penalty 4
Parameter fctp3(i,j) time cost for walking;
   fctp3(i,j)=f*d(i,j)+8*tpf(j);

Parameter bctp3(i,j) time cost for biking;
   bctp3(i,j)=b*d(i,j)+8*tpb(j);

Parameter cctp3(i,j) time cost for driving;
   cctp3(i,j)=c*d(i,j)+8*tpc(j);

** time penalty 5
Parameter fctp4(i,j) time cost for walking;
   fctp4(i,j)=f*d(i,j)+16*tpf(j);

Parameter bctp4(i,j) time cost for biking;
   bctp4(i,j)=b*d(i,j)+16*tpb(j);

Parameter cctp4(i,j) time cost for driving;
   cctp4(i,j)=c*d(i,j)+16*tpc(j);

** time penalty 6
Parameter fctp5(i,j) time cost for walking;
   fctp5(i,j)=f*d(i,j)+32*tpf(j);

Parameter bctp5(i,j) time cost for biking;
   bctp5(i,j)=b*d(i,j)+32*tpb(j);

Parameter cctp5(i,j) time cost for driving;
   cctp5(i,j)=c*d(i,j)+32*tpc(j);

Variables
     z
** vars with no time penalty
     xf(i,j)  number of people taking the route from i to j by foot
     xb(i,j)  number of people taking the route from i to j by bike
     xc(i,j)  number of people taking the route from i to j by car

** vars with time penalty 1
    xftp(i,j)  number of people taking the route from i to j by foot with tp
    xbtp(i,j)  number of people taking the route from i to j by bike with tp
    xctp(i,j)  number of people taking the route from i to j by car with tp

** vars with time penalty 2
    xftp1(i,j)
    xbtp1(i,j)
    xctp1(i,j)

** vars with time penalty 3
    xftp2(i,j)
    xbtp2(i,j)
    xctp2(i,j)

** vars with time penalty 4
    xftp3(i,j)
    xbtp3(i,j)
    xctp3(i,j)

** vars with time penalty 5
    xftp4(i,j)
    xbtp4(i,j)
    xctp4(i,j)

** vars with time penalty 6
    xftp5(i,j)
    xbtp5(i,j)
    xctp5(i,j);

positive variables
    xf
    xb
    xc

    xftp
    xbtp
    xctp

    xftp1
    xbtp1
    xctp1

    xftp2
    xbtp2

xctp2

xftp3
xbtp3
xctp3

xftp4
xbtp4
xctp4

xftp5
xbtp5
xctp5;

Equations
obj

**constraints
*total population must be fully evacuated
population(i)

*foot/bike/car traffic must be <= max capacities of each exit
maxcapfoot(j)
maxcapbike(j)
maxcapcar(j)

maxcapfoot0(j)
maxcapbike0(j)
maxcapcar0(j)

maxcapfoot1(j)
maxcapbike1(j)
maxcapcar1(j)

maxcapfoot2(j)
maxcapbike2(j)
maxcapcar2(j)

maxcapfoot3(j)
maxcapbike3(j)
maxcapcar3(j)

maxcapfoot4(j)
maxcapbike4(j)
maxcapcar4(j)

*combined foot and bike traffic must be <= ferry capacities
    maxcapferry_me
    maxcapferry_sc
    maxcapferry_ch
    maxcapferry_ws
    maxcapferry_wh

    maxcapferry_me0
    maxcapferry_sc0
    maxcapferry_ch0
    maxcapferry_ws0
    maxcapferry_wh0

    maxcapferry_me1
    maxcapferry_sc1
    maxcapferry_ch1
    maxcapferry_ws1
    maxcapferry_wh1

    maxcapferry_me2
    maxcapferry_sc2
    maxcapferry_ch2
    maxcapferry_ws2
    maxcapferry_wh2

    maxcapferry_me3
    maxcapferry_sc3
    maxcapferry_ch3
    maxcapferry_ws3
    maxcapferry_wh3

    maxcapferry_me4
    maxcapferry_sc4
    maxcapferry_ch4
    maxcapferry_ws4
    maxcapferry_wh4

*infeasible routes to tunnel (i.e., walking, biking)
    impossible_tunnel

*infeasible routes to ferry (i.e., car)
    impossible_ferry

*infeasible routes to train (i.e., car)
    impossible_train

*portion of evacuees that can use bikes
    bikes(i)

*portion of evacuees that can use cars
    cars(i);

obj..          z =e= sum((i, j), fc(i,j)*xf(i,j)+bc(i,j)*xb(i,j)+cc(i,j)*xc(i,j)
               +fctp(i,j)*xftp(i,j)+bctp(i,j)*xbtp(i,j)+cctp(i,j)*xctp(i,j)
               +fctp1(i,j)*xftp1(i,j)+bctp1(i,j)*xbtp1(i,j)+cctp1(i,j)*xctp1(i,j)
               +fctp2(i,j)*xftp2(i,j)+bctp2(i,j)*xbtp2(i,j)+cctp2(i,j)*xctp2(i,j)
               +fctp3(i,j)*xftp3(i,j)+bctp3(i,j)*xbtp3(i,j)+cctp3(i,j)*xctp3(i,j)
               +fctp4(i,j)*xftp4(i,j)+bctp4(i,j)*xbtp4(i,j)+cctp4(i,j)*xctp4(i,j)
               +fctp5(i,j)*xftp5(i,j)+bctp5(i,j)*xbtp5(i,j)+cctp5(i,j)*xctp5(i,j));

population(i)..   sum(j, xf(i,j)+xb(i,j)+xc(i,j)
                  +xftp(i,j)+xbtp(i,j)+xctp(i,j)
                  +xftp1(i,j)+xbtp1(i,j)+xctp1(i,j)
                  +xftp2(i,j)+xbtp2(i,j)+xctp2(i,j)
                  +xftp3(i,j)+xbtp3(i,j)+xctp3(i,j)
                  +xftp4(i,j)+xbtp4(i,j)+xctp4(i,j)
                  +xftp5(i,j)+xbtp5(i,j)+xctp5(i,j))=e=p(i);

maxcapfoot(j)..     sum(i, xf(i,j)) =l= mf(j);
maxcapbike(j)..     sum(i, xb(i,j)) =l= mb(j);
maxcapcar(j)..      sum(i, xc(i,j)) =l= mc(j);

maxcapfoot0(j)..    sum(i, xftp(i,j)) =l= mf(j);
maxcapbike0(j)..    sum(i, xbtp(i,j)) =l= mb(j);
maxcapcar0(j)..     sum(i, xctp(i,j)) =l= mc(j);

maxcapfoot1(j)..    sum(i, xftp1(i,j)) =l= mf(j);
maxcapbike1(j)..    sum(i, xbtp1(i,j)) =l= mb(j);
maxcapcar1(j)..     sum(i, xctp1(i,j)) =l= mc(j);

maxcapfoot2(j)..    sum(i, xftp2(i,j)) =l= mf(j);
maxcapbike2(j)..    sum(i, xbtp2(i,j)) =l= mb(j);
maxcapcar2(j)..     sum(i, xctp2(i,j)) =l= mc(j);

maxcapfoot3(j)..    sum(i, xftp3(i,j)) =l= mf(j);
maxcapbike3(j)..    sum(i, xbtp3(i,j)) =l= mb(j);
maxcapcar3(j)..     sum(i, xctp3(i,j)) =l= mc(j);

maxcapfoot4(j)..    sum(i, xftp4(i,j)) =l= mf(j);
maxcapbike4(j)..    sum(i, xbtp4(i,j)) =l= mb(j);
maxcapcar4(j)..     sum(i, xctp4(i,j)) =l= mc(j);

maxcapferry_me..    sum(i, xf(i,"midtown-east")+xb(i,"midtown-east")) =l= mf("midtown-east");
maxcapferry_sc..    sum(i, xf(i,"stuyvesant-cove")+xb(i,"stuyvesant-cove")) =l= mf("Stuyvesant-cove");
maxcapferry_ch..    sum(i, xf(i,"corlears-hook")+xb(i,"corlears-hook")) =l= mf("corlears-hook");
maxcapferry_ws..    sum(i, xf(i,"wall-street")+xb(i,"wall-street")) =l= mf("wall-street");
maxcapferry_wh..    sum(i, xf(i,"whitehall")+xb(i,"whitehall")) =l= mf("whitehall");

maxcapferry_me0..   sum(i, xftp(i,"midtown-east")+xbtp(i,"midtown-east")) =l= mf("midtown-east");
maxcapferry_sc0..   sum(i, xftp(i,"stuyvesant-cove")+xbtp(i,"stuyvesant-cove")) =l= mf("Stuyvesant-cove");
maxcapferry_ch0..   sum(i, xftp(i,"corlears-hook")+xbtp(i,"corlears-hook")) =l= mf("corlears-hook");
maxcapferry_ws0..   sum(i, xftp(i,"wall-street")+xbtp(i,"wall-street")) =l= mf("wall-street");
maxcapferry_wh0..   sum(i, xftp(i,"whitehall")+xbtp(i,"whitehall")) =l= mf("whitehall");

maxcapferry_me1..   sum(i, xftp1(i,"midtown-east")+xbtp1(i,"midtown-east")) =l= mf("midtown-east");
maxcapferry_sc1..   sum(i, xftp1(i,"stuyvesant-cove")+xbtp1(i,"stuyvesant-cove")) =l= mf("Stuyvesant-cove");
maxcapferry_ch1..   sum(i, xftp1(i,"corlears-hook")+xbtp1(i,"corlears-hook")) =l= mf("corlears-hook");
maxcapferry_ws1..   sum(i, xftp1(i,"wall-street")+xbtp1(i,"wall-street")) =l= mf("wall-street");
maxcapferry_wh1..   sum(i, xftp1(i,"whitehall")+xbtp1(i,"whitehall")) =l= mf("whitehall");

maxcapferry_me2..   sum(i, xftp2(i,"midtown-east")+xbtp2(i,"midtown-east")) =l= mf("midtown-east");
maxcapferry_sc2..   sum(i, xftp2(i,"stuyvesant-cove")+xbtp2(i,"stuyvesant-cove")) =l= mf("Stuyvesant-cove");
maxcapferry_ch2..   sum(i, xftp2(i,"corlears-hook")+xbtp2(i,"corlears-hook")) =l= mf("corlears-hook");
maxcapferry_ws2..   sum(i, xftp2(i,"wall-street")+xbtp2(i,"wall-street")) =l= mf("wall-street");
maxcapferry_wh2..   sum(i, xftp2(i,"whitehall")+xbtp2(i,"whitehall")) =l= mf("whitehall");

maxcapferry_me3..   sum(i, xftp3(i,"midtown-east")+xbtp3(i,"midtown-east")) =l= mf("midtown-east");
maxcapferry_sc3..   sum(i, xftp3(i,"stuyvesant-cove")+xbtp3(i,"stuyvesant-cove")) =l= mf("Stuyvesant-cove");

maxcapferry_ch3..     sum(i, xftp3(i,"corlears-hook")+xbtp3(i,"corlears-hook")) =l= mf("corlears-hook");
maxcapferry_ws3..     sum(i, xftp3(i,"wall-street")+xbtp3(i,"wall-street")) =l= mf("wall-street");
maxcapferry_wh3..     sum(i, xftp3(i,"whitehall")+xbtp3(i,"whitehall")) =l= mf("whitehall");

maxcapferry_me4..     sum(i, xftp4(i,"midtown-east")+xbtp4(i,"midtown-east")) =l= mf("midtown-east");
maxcapferry_sc4..     sum(i, xftp4(i,"stuyvesant-cove")+xbtp4(i,"stuyvesant-cove")) =l= mf("Stuyvesant-cove");
maxcapferry_ch4..     sum(i, xftp4(i,"corlears-hook")+xbtp4(i,"corlears-hook")) =l= mf("corlears-hook");
maxcapferry_ws4..     sum(i, xftp4(i,"wall-street")+xbtp4(i,"wall-street")) =l= mf("wall-street");
maxcapferry_wh4..     sum(i, xftp4(i,"whitehall")+xbtp4(i,"whitehall")) =l= mf("whitehall");

impossible_tunnel..    sum(i, xf(i,"queens-midtown-tunnel")+xb(i,"queens-midtown-tunnel")+xftp(i,"queens-midtown-tunnel")+xbtp(i,"queens-midtown-tunnel")+xftp1(i,"queens-midtown-tunnel")+xbtp1(i,"queens-midtown-tunnel")
              +xftp2(i,"queens-midtown-tunnel")+xbtp2(i,"queens-midtown-tunnel")+xftp3(i,"queens-midtown-tunnel")+xbtp3(i,"queens-midtown-tunnel")+xftp4(i,"queens-midtown-tunnel")+xbtp4(i,"queens-midtown-tunnel")+xftp5(i,"queens-midtown-tunnel")+xbtp5(i,"queens-midtown-tunnel")
              +xf(i,"lincoln-tunnel")+xb(i,"lincoln-tunnel")+xftp(i,"lincoln-tunnel")+xbtp(i,"lincoln-tunnel")+xftp1(i,"lincoln-tunnel")+xbtp1(i,"lincoln-tunnel")
              +xftp2(i,"lincoln-tunnel")+xbtp2(i,"lincoln-tunnel")+xftp3(i,"lincoln-tunnel")+xbtp3(i,"lincoln-tunnel")+xftp4(i,"lincoln-tunnel")+xbtp4(i,"lincoln-tunnel")+xftp5(i,"lincoln-tunnel")+xbtp5(i,"lincoln-tunnel")
              +xf(i,"holland-tunnel")+xb(i,"holland-tunnel")+xftp(i,"holland-tunnel")+xbtp(i,"holland-tunnel")+xftp1(i,"holland-tunnel")+xbtp1(i,"holland-tunnel")
              +xftp2(i,"holland-tunnel")+xbtp2(i,"holland-tunnel")+xftp3(i,"holland-tunnel")+xbtp3(i,"holland-tunnel")+xftp4(i,"holland-tunnel")+xbtp4(i,"holland-tunnel")+xftp5(i,"holland-tunnel")+xbtp5(i,"holland-tunnel")
              +xf(i,"hlc-tunnel")+xb(i,"hlc-tunnel")+xftp(i,"hlc-tunnel")+xbtp(i,"hlc-tunnel")+xftp1(i,"hlc-tunnel")+xbtp1(i,"hlc-tunnel")
              +xftp2(i,"hlc-tunnel")+xbtp2(i,"hlc-tunnel")+xftp3(i,"hlc-tunnel")+xbtp3(i,"hlc-tunnel")+xftp4(i,"hlc-tunnel")+xbtp4(i,"hlc-tunnel")+xftp5(i,"hlc-tunnel")+xbtp5(i,"hlc-tunnel")) =e= 0;

impossible_ferry..    sum(i, xc(i,"midtown-east")+xctp(i,"midtown-east")+xctp1(i,"midtown-east")+xctp2(i,"midtown-east")+xctp3(i,"midtown-east")+xctp4(i,"midtown-east")+xctp5(i,"midtown-east")
              +xc(i,"stuyvesant-cove")+xctp(i,"stuyvesant-cove")+xctp1(i,"stuyvesant-cove")+xctp2(i,"stuyvesant-cove")+xctp3(i,"stuyvesant-cove")+xctp4(i,"stuyvesant-cove")+xctp5(i,"stuyvesant-cove")

```
                +xc(i,"corlears-hook")+xctp(i,"corlears-hook")+xctp1(i,"corlears-
hook")+xctp2(i,"corlears-hook")+xctp3(i,"corlears-hook")+xctp4(i,"corlears-
hook")+xctp5(i,"corlears-hook")
                +xc(i,"wall-street")+xctp(i,"wall-street")+xctp1(i,"wall-street")+xctp2(i,"wall-
street")+xctp3(i,"wall-street")+xctp4(i,"wall-street")+xctp5(i,"wall-street")

+xc(i,"whitehall")+xctp(i,"whitehall")+xctp1(i,"whitehall")+xctp2(i,"whitehall")+xctp3(i,"white
hall")+xctp4(i,"whitehall")+xctp5(i,"whitehall"))  =e= 0;

impossible_train..    sum(i, xc(i,"penn-station")+xctp(i,"penn-station")+xctp1(i,"penn-
station")+xctp2(i,"penn-station")+xctp3(i,"penn-station")+xctp4(i,"penn-station")+xctp5(i,"penn-
station"))  =e= 0;

bikes(i)..       sum(j, xb(i,j)+xbtp(i,j)+xbtp1(i,j)+xbtp2(i,j)+xbtp3(i,j)+xbtp4(i,j)+xbtp5(i,j))
=l= p(i)*.24;

cars(i)..        sum(j, xc(i,j)+xctp(i,j)+xctp1(i,j)+xctp2(i,j)+xctp3(i,j)+xctp4(i,j)+xctp5(i,j))
=l= p(i)*.22;

Model evacuation_model / all /;
solve evacuation_model using lp minimizing z;

** display
display z.l ;
display xf.l;
display xb.l;
display xc.l;
display xftp.l;
display xbtp.l;
display xctp.l;
display xftp1.l;
display xbtp1.l;
display xctp1.l;
display xftp2.l;
display xbtp2.l;
display xctp2.l;
display xftp3.l;
display xbtp3.l;
display xctp3.l;
display xftp4.l;
```

```
display xbtp4.l;
display xctp4.l;
display xftp5.l;
display xbtp5.l;
display xctp5.l;
display fc;
display bc;
display fctp4;
display bctp4;
```

****************** total **********************
```
display "***************total*******************"
Parameter tot_foot;
   tot_foot= sum((i,j),
xf.L(i,j)+xftp.L(i,j)+xftp1.L(i,j)+xftp2.L(i,j)+xftp3.L(i,j)+xftp4.L(i,j)+xftp5.L(i,j));
display tot_foot;

Parameter tot_bikes;
   tot_bikes= sum((i,j),
xb.L(i,j)+xbtp.L(i,j)+xbtp1.L(i,j)+xbtp2.L(i,j)+xbtp3.L(i,j)+xbtp4.L(i,j)+xbtp5.L(i,j));
display tot_bikes;

Parameter tot_cars;
   tot_cars= sum((i,j),
xc.L(i,j)+xctp.L(i,j)+xctp1.L(i,j)+xctp2.L(i,j)+xctp3.L(i,j)+xctp4.L(i,j)+xctp5.L(i,j));
display tot_cars;
```

***************** east-village *********************
```
display "***************east-village*******************"
Parameter mw_foot;
mw_foot = sum(j,xf.L("east-village",j)+xftp.L("east-village",j)+xftp1.L("east-village",j)+xftp2.L("east-village",j)+xftp3.L("east-village",j)+xftp4.L("east-village",j)+xftp5.L("east-village",j));
display mw_foot;

Parameter mw_bike;
   mw_bike = sum(j,xb.L("east-village",j)+xbtp.L("east-village",j)+xbtp1.L("east-village",j)+xbtp2.L("east-village",j)+xbtp3.L("east-village",j)+xbtp4.L("east-village",j)+xbtp5.L("east-village",j));
display mw_bike;

Parameter mw_cars;
   mw_cars = sum(j,xc.L("east-village",j)+xctp.L("east-village",j)+xctp1.L("east-village",j)+xctp2.L("east-village",j)+xctp3.L("east-village",j)+xctp4.L("east-village",j)+xctp5.L("east-village",j));
display mw_cars;
```

****************** midtown-west **********************
display "***************midtown-west*******************"
Parameter mw_foot;
mw_foot = sum(j,xf.L("midtown-west",j)+xftp.L("midtown-west",j)+xftp1.L("midtown-west",j)+xftp2.L("midtown-west",j)+xftp3.L("midtown-west",j)+xftp4.L("midtown-west",j)+xftp5.L("midtown-west",j));
display mw_foot;

Parameter mw_bike;
   mw_bike = sum(j,xb.L("midtown-west",j)+xbtp.L("midtown-west",j)+xbtp1.L("midtown-west",j)+xbtp2.L("midtown-west",j)+xbtp3.L("midtown-west",j)+xbtp4.L("midtown-west",j)+xbtp5.L("midtown-west",j));
display mw_bike;

Parameter mw_cars;
   mw_cars = sum(j,xc.L("midtown-west",j)+xctp.L("midtown-west",j)+xctp1.L("midtown-west",j)+xctp2.L("midtown-west",j)+xctp3.L("midtown-west",j)+xctp4.L("midtown-west",j)+xctp5.L("midtown-west",j));
display mw_cars;

****************** midtown-east **********************
display "***************midtown-east*******************"
Parameter me_foot;
me_foot = sum(j,xf.L("midtown-east",j)+xftp.L("midtown-east",j)+xftp1.L("midtown-east",j)+xftp2.L("midtown-east",j)+xftp3.L("midtown-east",j)+xftp4.L("midtown-east",j)+xftp5.L("midtown-east",j));
display me_foot;

Parameter me_bike;
  me_bike = sum(j,xb.L("midtown-east",j)+xbtp.L("midtown-east",j)+xbtp1.L("midtown-east",j)+xbtp2.L("midtown-east",j)+xbtp3.L("midtown-east",j)+xbtp4.L("midtown-east",j)+xbtp5.L("midtown-east",j));
display me_bike;

Parameter me_cars;
  me_cars = sum(j,xc.L("midtown-east",j)+xctp.L("midtown-east",j)+xctp1.L("midtown-east",j)+xctp2.L("midtown-east",j)+xctp3.L("midtown-east",j)+xctp4.L("midtown-east",j)+xctp5.L("midtown-east",j));
display me_cars;

*** display maximums ***
**
Parameter max_fc;
   max_fc = smax((i,j)$(xf.L(i,j)>0), fc(i,j));

```
*    display max_fc;

Parameter max_bc;
   max_bc = smax((i,j)$(xb.L(i,j)>0), bc(i,j));
*    display max_bc;

Parameter max_cc;
   max_cc = smax((i,j)$(xc.L(i,j)>0), cc(i,j));
*    display max_cc;
**
Parameter max_fctp;
   max_fctp = smax((i,j)$(xftp.L(i,j)>0), fctp(i,j));
*    display max_fctp;

Parameter max_bctp;
   max_bctp = smax((i,j)$(xbtp.L(i,j)>0), bctp(i,j));
*    display max_bctp;

Parameter max_cctp;
   max_cctp = smax((i,j)$(xctp.L(i,j)>0), cctp(i,j));
*    display max_cctp;

**
Parameter max_fctp1;
   max_fctp1 = smax((i,j)$(xftp1.L(i,j)>0), fctp1(i,j));
*    display max_fctp1;

Parameter max_bctp1;
   max_bctp1 = smax((i,j)$(xbtp1.L(i,j)>0), bctp1(i,j));
*    display max_bctp1;

Parameter max_cctp1;
   max_cctp1 = smax((i,j)$(xctp1.L(i,j)>0), cctp1(i,j));
*    display max_cctp1;
**
Parameter max_fctp2;
   max_fctp2 = smax((i,j)$(xftp2.L(i,j)>0), fctp2(i,j));
*    display max_fctp2;

Parameter max_bctp2;
   max_bctp2 = smax((i,j)$(xbtp2.L(i,j)>0), bctp2(i,j));
*    display max_bctp2;

Parameter max_cctp2;
   max_cctp2 = smax((i,j)$(xctp2.L(i,j)>0), cctp2(i,j));
*    display max_cctp2;
```

**

Parameter max_fctp3;
   max_fctp3 = smax((i,j)$(xftp3.L(i,j)>0), fctp3(i,j));
*   display max_fctp3;

Parameter max_bctp3;
   max_bctp3 = smax((i,j)$(xbtp3.L(i,j)>0), bctp3(i,j));
*   display max_bctp3;

Parameter max_cctp3;
   max_cctp3 = smax((i,j)$(xctp3.L(i,j)>0), cctp3(i,j));
*   display max_cctp3;

**

Parameter max_fctp4;
   max_fctp4 = smax((i,j)$(xftp4.L(i,j)>0), fctp4(i,j));
*   display max_fctp4;

Parameter max_bctp4;
   max_bctp4 = smax((i,j)$(xbtp4.L(i,j)>0), bctp4(i,j));
*   display max_bctp4;

Parameter max_cctp4;
   max_cctp4 = smax((i,j)$(xctp4.L(i,j)>0), cctp4(i,j));
*   display max_cctp4;

**

Parameter max_fctp5;
   max_fctp5 = smax((i,j)$(xftp5.L(i,j)>0), fctp5(i,j));
*   display max_fctp5;

Parameter max_bctp5;
   max_bctp5 = smax((i,j)$(xbtp5.L(i,j)>0), bctp5(i,j));
*   display max_bctp5;

Parameter max_cctp5;
   max_cctp5 = smax((i,j)$(xctp5.L(i,j)>0), cctp5(i,j));
*   display max_cctp5;

** display total evacuation time
Parameter totalevac_in_hrs;
   totalevac_in_hrs = max(max_fc, max_bc, max_cc,
            max_fctp, max_fctp, max_fctp,
            max_fctp1, max_fctp1, max_fctp1,
            max_fctp2, max_fctp2, max_fctp2,
            max_fctp3, max_fctp3, max_fctp3,

```
            max_fctp4, max_fctp4, max_fctp4,
            max_fctp5, max_fctp5, max_fctp5)/60;
   display totalevac_in_hrs;
   display evacuation_model.numvar ;

*=== Export to Excel using GDX utilities

*=== First unload to GDX file (occurs during execution phase)
execute_unload "results1.gdx" z.L xf.l xb.l xc.l xftp.l xbtp.l xctp.l xftp1.l xbtp1.l xctp1.l xftp2.l
xbtp2.l xctp2.l xftp3.l xbtp3.l xctp3.l xftp4.l xbtp4.l xctp4.l xftp5.l xbtp5.l xbtp5.l xctp5.l

*=== Now write to variable levels to Excel file from GDX
*=== Since we do not specify a sheet, data is placed in first sheet
execute 'gdxxrw.exe results1.gdx o=results1.xls var=xf.L'
```